\documentclass{sigchi}

\usepackage{balance}  
\usepackage{color}  
\usepackage{graphics} 
\usepackage{ifpdf}
\usepackage{times}    
\usepackage[hyphens]{url}      
\usepackage{enumitem}
\usepackage[figuresleft]{rotating}
\usepackage{array,booktabs}

\toappear{}

\makeatletter
\def\url@leostyle{%
  \@ifundefined{selectfont}{\def\UrlFont{\sf}}{\def\UrlFont{\small\bf\ttfamily}}}
\makeatother
\ifpdf
\def\pprw{8.5in}
\def\pprh{11in}

\fi

\ifpdf
\usepackage[pdftex]{hyperref}
\hypersetup{
pdftitle={SIGCHI Conference Proceedings Format},
pdfauthor={LaTeX},
pdfkeywords={SIGCHI, proceedings, archival format},
pdfstartview={FitH},
bookmarksnumbered,
colorlinks,
citecolor=black,
filecolor=black,
linkcolor=black,
urlcolor=black,
breaklinks=true,
}
\fi

\begin{document}

\title{Beyond AMT: An Analysis of Crowd Work Platforms\\~}


\numberofauthors{2}
\author{
  \alignauthor Donna Vakharia\\
    \affaddr{Department of Computer Science}\\
    \affaddr{University of Texas at Austin}\\
    \email{donna@cs.utexas.edu}
  \alignauthor Matthew Lease\\
    \affaddr{School of Information}\\
    \affaddr{University of Texas at Austin}\\
    \email{ml@ischool.utexas.edu}
\vspace{10pt}
}
%
%


\maketitle

\begin{abstract}
While Amazon's Mechanical Turk (AMT) helped launch the paid {\em crowd work} industry eight years ago, 
many new vendors now offer a range of alternative models. 
Despite this, little crowd work research has 
explored other platforms. 
Such near-exclusive focus risks letting AMT's particular vagaries and limitations overly shape our understanding of crowd work and the research questions and directions being pursued. 
To address this, we present a cross-platform content analysis of seven crowd work platforms. We begin by reviewing how AMT assumptions and limitations have influenced prior research. Next, we formulate key criteria for characterizing and differentiating crowd work platforms. Our analysis of platforms contrasts them with AMT, informing both methodology of use and directions for future research. Our cross-platform analysis represents the only such study by researchers for researchers, intended to  
further enrich the diversity of research on crowd work and accelerate progress.
\end{abstract}

%

%
%
%
%


\section{Introduction}


Amazon's Mechanical Turk (AMT)~\cite{barr2006ai,Mieszkowski06,chen2011opportunities} has revolutionized data processing and collection practice in both research and industry, and it remains one of the most prominent paid {\em crowd work}~\cite{Kittur13} platforms today. 
Unfortunately, it remains in beta eight years after its launch 
with many of the same limitations as then:
lack of worker profiles indicating skills or experience, inability to post worker or employer ratings and reviews, minimal infrastructure for effectively managing workers or collecting analytics, etc. 
Achieving quality, complex work with AMT remains challenging.

Since AMT 
helped launch the crowd work industry eight years ago, 
a myriad of new vendors have arisen which now offer a wide range of features and workflow models for accomplishing quality work (\url{crowdsortium.org}). Nonetheless, research on crowd work has continued to focus on AMT near-exclusively. 
By analogy, if one's only experience of programming came from using {\tt Basic}, how might this 
limit one's overall conception of programming? What if the only search engine we knew was {\it AltaVista}? Adar opined that prior research has often been envisioned too narrowly for AMT, ``...writing the user's manual for MTurk ... struggl[ing] against the limits of the platform...''~\cite{adar2011hate}. Such focus risks letting AMT's particular vagaries and limitations unduly shape our research questions, methodology, and imagination.

To assess the extent of AMT's influence upon research questions and use, we review its impact on prior work, assess what functionality and workflows other platforms offer, and consider what light other platforms' diverse capabilities sheds on current research practices and future directions. 
To this end, we present a comparative content analysis~\cite{mayring2000qualitative} of seven alternative crowd work platforms: ClickWorker, CloudFactory, CrowdComputing Systems, 
CrowdFlower, CrowdSource, MobileWorks, and oDesk. 

To characterize and differentiate crowd work platforms, we identify several key categories for analysis. Our content analysis assesses each platform by drawing upon a variety of information sources: Webpages, blogs, news articles, white papers, and research papers. We also shared our analysis of each platform with its representatives and requested their feedback, which we have incorporated. We focus our first cross-platform study on a qualitative analysis, leaving a quantitative evaluation for future work. We expect each approach to complement the other and yield distinct insights.


{\bf Contributions.} 
Our content analysis of crowd work platforms represents the first such study we know of by researchers for researchers, with categories of analysis chosen based on research relevance. Contributions include our review of how AMT assumptions and limitations have influenced prior research, the detailed criteria we developed for characterizing crowd work platforms, and our analysis. 
Findings inform both methodology for use of today's platforms and discussion of future research directions. 
Our study is expected to foster more crowd work research beyond AMT, 
further enriching research diversity and accelerating progress.

While our detailed description of current platform capabilities may usefully guide platform selection today, we expect these details to become quickly dated as the industry continues to rapidly evolve. Instead, we expect a more enduring contribution of our work is simply illustrating a wide range of current designs, provoking more exploration beyond use of AMT and more research studies that defy closely-coupling to AMT's particular design. 
A retrospective contribution of our work may be its snapshot in time of today's crowd platforms, providing a baseline or inspiration for future designs.

\section{Related Work}

Human computation and crowdsourcing~\cite{Quinn11,Grier05,Law11,Davis10,Howe06} arise in many forms, including gamification~\cite{von2004labeling}, citizen science~\cite{evans2007archives}, peer production/co-creation~\cite{viegas2007hidden}, wisdom of crowds~\cite{Surowiecki05}, and collective intelligence~\cite{ci2012}, among others. In this paper, we focus on paid {\em crowd work}~\cite{Kittur13}, such as on AMT~\cite{barr2006ai,Mieszkowski06,paolacci2010running} where {\em Requesters} post paid tasks to be completed by {\em Workers}. 
For researchers interested in data collection or integrating human computation into larger software systems, AMT research has been prolific, especially in the areas of computer vision~\cite{sorokin2008utility}, human-computer interaction (HCI)~\cite{Kittur08}, natural language processing (NLP)~\cite{Snow08}, information retrieval (IR)~\cite{Alonso08}, and behavioral science~\cite{buhrmester2011amazon,mason2012conducting}. 



Discussion of AMT limitations has a long history~\cite{atwood2007amazon,Mims10}, including criticism as a ``market for lemons''~\cite{Ipeirotis-lemons} and ``digital sweatshop''~\cite{Cushing13}. In 2010, Ipeirotis~\cite{panos2010pleatoamazon} called for: a better interface to post tasks, a better worker reputation system~\cite{ipeirotis-be-a-top-worker}, a requester trustworthiness guarantee, and a better task search interface for workers~\cite{chilton2010task}. Many studies note quality concerns and suggest ways to address them~\cite{Ipeirotis:Wang:2010,dow2012shepherding}. Relatively few safeguards protect against fraud; while terms-of-use prohibit robots and multiple account use by a single worker ({\em sybil attacks}~\cite{levine2006survey}), anecdotal evidence suggests lax enforcement~\cite{McCreadie-CSE2010,zhu10}. While such {\em spammer} fraud is oft-discussed~\cite{Kittur08,downs2010your,heymann2011turkalytics,difallah2012mechanical,raykar2012eliminating,Eickhoff13}, fraud by Requesters is also problematic~\cite{panos2010pleatoamazon,tn-avoid}. Little support exists to price tasks for effectiveness or fairness~\cite{Mason:Watts:2009,faridani2011s,singer2011pricing}. 

With no native support for workflows or collaboration, the challenge of tackling complex work has led to new methodology~\cite{kittur2012crowdweaver,kulkarni2012collaboratively,ahmad2011jabberwocky,noronha2011platemate} and open-source toolkits~\cite{little2009turkit,kittur2011crowdforge} (e.g., \url{code.google.com/p/quikturkit} and \url{twitter.github.io/clockworkraven}). 
AMT provides no native support for hierarchical management structures, a hallmark of traditional organizational practice~\cite{kochhar2010anatomy}. No support is provided for routing tasks or examples to the most appropriate workers~\cite{Ho2012}, and as noted earlier, it can be difficult for workers to find tasks of interest~\cite{chilton2010task,law2011effects}. How to perform near real-time work has attracted much attention~\cite{bigham2010vizwiz,bernstein2010soylent,bernstein2011crowds,lasecki2012real}. 

A less discussed limitation of AMT is the absence of support for ``private'' crowds, differentiating efficiency of a crowdsourcing workflow vs.\ which workforce is actually performing the work.  In particular, sensitive data may be subject to federal or state regulations (e.g., customer or student data) or have other privacy conerns (e.g., intellectual property, trade secrets, intelligence and security, etc.). A closed crowd may have signed non-disclosure agreements (NDAs) providing a legal guarantees of safeguarding requester data~\cite{Nallapati13}. 

Lack of worker analytics has also led to various methods being proposed~\cite{heymann2011turkalytics,Rzeszotarski11}, while lack of access to worker demographics has led various researchers to collect such data~\cite{Ross10,ipeirotis2010demographics}. Inability of workers' to share identity and skills has prompted some to integrate crowd platforms with social networks~\cite{difallah13}. Tasks requiring foreign language skills require requesters to design their own quality tests to check the competence~\cite{mellebeek2010opinion}, and AMT may even be no longer allowing international workers to join~\cite{no-intl}. For certain tasks, such as human subjects research~\cite{mason2012conducting}, undesired access to worker identities can complicate research oversight. For other tasks, knowledge of worker identity can inform credibility assessment of work products, as well as provide greater confidence that work is being completed by a real person and not a script bot. 

Quinn and Bederson~\cite{Quinn11} provide a very useful literature review of human computation, organized around the dimensions of motivation, quality control, aggregation, human skill, process order, and task-request cardinality. While these dimensions provide a useful organizing framework for prior work, the dimensions are related but not well-matched for assessing platforms. The authors briefly mention ChaCha, LiveOps, and CrowdFlower platforms, but most of their platform examples cite AMT, reflecting prior work's focus.

Kittur et al.~\cite{Kittur13} provide a more applicable conceptual organization of crowd work research areas: workflow, motivation, hierarchy, reputation, collaboration, real-time work, quality assurance, career ladders, communication channels, and learning opportunities. 
While they did not address current platforms' capabilities and limitations, they identify important future research, and their conceptual areas inspired our initial categories for analyzing crowd work platforms. 

\vspace{-5pt} 
\subsection{Other Platforms \& Comparisons}

CrowdConf 2010 (\url{www.crowdconf2010.com}) helped increase researcher awareness of AMT alternatives~\cite{ipeirotis-micro-services,ipeirotis-verticals}. That same year, CrowdFlower~\cite{Le10,Oleson11} and AMT co-sponsored an NLP workshop~\cite{Burch-amt10}. One of the few papers we know of contrasting results across platforms stems from this workshop. 

Finin et al.~\cite{finin2010annotating} contrast AMT vs.\ CrowdFlower for named entity annotation of Twitter status updates. Use of standard HTML and Javascript vs.\ CrowdFlower�s CrowdFlower Mark-up Language (CML) was reported as an AMT advantage. For CrowdFlower, support for data verification via built-in gold standard tests was valued, along with allowing work across multiple workforce channels, ability to get more judgments to improve quality in cases as needed, pricing assistance, workers getting immediate feedback when missing gold tests, access to detail management \& analytic tools for overseeing workers, a calibration interface that assists in deciding pay rates, and automatic pausing of HITs in case of high workers' error rates on gold tests. However, they were able to duplicate some of the gold standard functionality on AMT by combining regular and quality control queries at the cost of losing the added functionality and convenience offered by CrowdFlower�s platform facilitated gold tests. 

In another paper from the workshop, Negri et al.~\cite{negri2010creating} noted CrowdFlower's lack of region-based and other qualification mechanisms supported by AMT. Of course, with the platforms and industry rapidly changing, we must expect some findings may become quickly dated.
Research papers by platform personnel are cited later when we assess platforms.

\vspace{-5pt} 
\subsection{Industrial White Papers}

We are not aware of any research studies performing systematic cross-platform analysis, qualitative or quantitative. However, several industrial white papers~\cite{frei2009paid,Turian12,info-evolution,CCSWhitepaper} 
provide a helpful starting point. 
NYU alumnus Joseph Turian follows Ipeirotis~\cite{ipeirotis-channel} in decomposing crowd work into a tripartite {\em stack} of workforce, a platform, and/or applications~\cite{Turian12}. Turian identifies the platform as the limiting component and least mature part of the stack today. 

Turian compared CrowdComputing Systems, CloudFactory, Servio, CrowdSource, CrowdFlower, and MobileWorks, but his report is more industry-oriented as it provides comparative analysis with traditional Business Process Outsourcing (BPO) approaches. Crowd labor, built on a modular stack, is noted as more efficient as compared to traditional BPO and is deflating this traditional BPO market. Also several industry-specific crowd labor applications such as social media management, retail analytics, granular sentiment analysis for ad agencies, and product merchandising are discussed.  Quality in particular is reported to be the key concern of personnel from every platform, with AMT being ``plagued by low-quality results.'' 
Besides Quality, and BPO, he identifies three other key ``disruption vectors'' for future work: 1) specialized labor (e.g., highly skilled, creative, and ongoing); 2)  application specialization ; 3) marketplaces and business models. These vectors are assigned scores (in some undefined way) based on relative importance. Platforms' comparative scores are computed as a function of these scores for each vector. 

A CrowdComputing Systems white paper~\cite{CCSWhitepaper} discusses how enterprise-crowdsourcing and crowd-computing can disrupt the outsourcing industry. It provides a matrix comparing crowdsourcing platforms (CrowdComputing Systems, AMT, Elance/oDesk, CrowdFlower), BPO and BPM (Business Process Management). The matrix provides binary results, as to whether the feature is supported or not. However, the selection criteria, explanation, and importance of the features used for comparison are not defined.

A white paper by Smartsheet~\cite{frei2009paid} compared 50 paid crowdsourcing vendors which was based on business applicability perspective, considering user experience (crowd responsiveness, ease of use, satisfactory results, cost advantage, private and secure) and infrastructure (crowd source, work definition and proposal, work and process, oversight, results and quality management, payment processing, API). While the resultant matrix showed the degree to which each platform supports comparison components, how these scores are calculated is unspecified, and no empirical comparison is reported.


\section{Criteria for Platform Assessments}
\label{sec:criteria}


This section defines the categories developed to characterize and differentiate crowd work platforms for our content analysis~\cite{mayring2000qualitative}. Criteria were inspired by Kittur et al.~\cite{Kittur13}, with inductive category development occurring during a first-pass, open-ended review of all platforms under consideration. Discussion after this first pass led to significant revision, followed by deductive application of categories. As boundary cases were encountered while coding (assessing) platforms according to criteria, cases were reviewed and prompted further revisions to category definitions to improve agreement.




{\bf Distinguishing Features.} Whereas other criteria are intentionally self-contained, {\em distinguishing features} summarize and contrast key platform features. What platform aspects particularly merit attention? 
A platform may provide access to workers in more regions of the world or otherwise differentiate its workforce. Advanced support might be offered for crowd work beyond the desktop, e.g., mobile SMS~\cite{eagle2009txteagle}, etc. 

{\bf Whose Crowd?} Does the platform maintain its own workforce, does it rely on other vendor ``channels'' to provide its workers, or is some hybrid combination of both labor sources adopted? 
Does the platform allow a requester to utilize and restrict tasks to his own private workforce or 
a closed crowd offering guarantees safeguarding of sensitive data~\cite{Nallapati13}?  
A Requester may want to exploit a platform's tool-suite for non-sensitive data but simply utilize his own known workforce. 
 
{\bf Demographics \& Worker Identities.} What demographic information is provided about the workforce~\cite{Ross10,ipeirotis2010demographics}?  How is this information made available to Requesters: individually or in aggregate?  Can Requesters specify desired/required demographics for a given task, and how easy and flexible is this? Are worker identities known to Requesters, or even the platform? What if any identify verification is performed? 

 
{\bf Qualifications \& Reputation.} Is language proficiency data provided?  Is some form of reputation tracking and/or skills listing associated with individual workers so that Requesters may better recruit, assess, and or manage workers? Are workers' interests or general expertise recorded? Is such data self-reported or assessed? How informative is whatever tracking system(s) used? How valid is information provided, and how robust is it to fraud and abuse~\cite{ipeirotis-be-a-top-worker}? 
What reputation or skill tracking systems are utilized (e.g., badges or certifications/qualifications, leader boards, tiered status levels, test scores, work history, ratings, comments, etc.). Are training sessions and/or competency tests offered or required?

{\bf Task Assignments \& Recommendations.} Is support provided for routing tasks or examples to the most appropriate workers~\cite{Ho2012}? How can workers effectively find tasks for which they are most interested and best suited~\cite{chilton2010task,law2011effects}? Are task assignments 
selected or suggested? 
How can Requesters find the best workers for different tasks? Does the platform detect and address task starvation 
to reduce task latency~\cite{dean2008mapreduce}?
 
{\bf Hierarchy \& Collaboration.} What support allows effective organization and coordination of workers, e.g. for traditional, hierarchical management structures~\cite{kochhar2010anatomy,Nallapati13}, or into teams for collaborative projects~\cite{anagnostopoulos2012online}? If peer review or assessment is utilized~\cite{horton2010employer}, how is it implemented?  How are alternative organizational structures determined and implemented across varying task types and complexity~\cite{noronha2011platemate,Heer10,lasecki2012real}? How does the platform facilitate effective communication and/or collaboration, especially as questions arise during work? 

{\bf Incentive Mechanisms.} 
What incentive mechanisms are offered to promote Worker participation (recruitment and retention) and effective work practices~\cite{shaw2011designing}? How are these incentives utilized individually or in combination? How are intrinsic vs.\ extrinsic rewards selected and managed for competing effects? How are appropriate incentives determined in accordance with the nature and constraints of varying tasks? 
 
{\bf Quality Assurance \& Control.} What quality assurance (QA) support is provided to ensure quality task design~\cite{huang2010toward}, and/or how are errors in submitted work detected/corrected via quality control (QC)~\cite{smyth1995inferring}? 
What do Requesters need to do and what is done for them?  What organizational structures and processes are utilized for QA and QC? For QA, how are Requesters enabled to design and deploy tasks to maximize result quality, e.g., providing task templates~\cite{chen2011opportunities}? What support is provided for the design of effective workflows~\cite{little2009turkit} and task 
interfaces? 
How are workers supported with appropriate tools, feedback, and assistance~\cite{dow2012shepherding}? For QC, what organizational or statistical techniques are applied to monitor workers and work products? Does the platform detect when a worker is struggling on a given task, and what support is provided if so? How is work aggregated to improve consensus quality?  
 
{\bf Self-service, Enterprise, and API offerings.} 
Enterprise ``white glove'' offerings are expected to provide high quality and may account for 50-90\% of platform revenue today 
\cite{Turian12}. 
Self-service solutions can be utilized directly by a Requester via the Web, typically without interaction with platform personnel. Does the platform provide a programmatic API for automating task management and integrating crowd work into software applications (for either self-service or enterprise solutions)? 
%
%
For enterprise-level solutions, how does the platform conceptualize the crowd work process, and who does what? 
How do different  offerings balance competing concerns of price vs.\ desired outcomes of work products?
 
{\bf Specialized \& Complex Task Support.} Are 
one or more vertical or horizontal niches of specialization offered as a particular strength, e.g.\ real-time transcription~\cite{lasecki2012real}? How does it innovate crowd work for such tasks? 
How does the platform enable tasks of increasing complexity to be effectively completed by 
crowds~\cite{noronha2011platemate,Heer10}?  
Does the platform offer ready task workflows or interfaces for these tasks, support for effective task decomposition and recombination, or design tools for creating such tasks more easily or effectively? Are specialized or complex tasks offered via a programmatic API? 
 

\begin{sidewaystable*}
\tiny
    \centering
    \caption{Assessment of Platforms}

\begin{tabular}[h]{| p{1.5cm}| p{2.7cm}| p{2.7cm}| p{2.7cm}| p{2.7cm}| p{2.7cm}| p{2.7cm}| p{2.7cm}|}
\hline
{\bf Features} & {\bf Clickworker} &{\bf CloudFactory} & {\bf CrowdComputing Systems} &{\bf CrowdFlower} &{\bf CrowdSource} &{\bf MobileWorks} &{\bf oDesk} \\ 
\hline

{\bf Distinguishing Features}  
&

{
\begin{itemize} [noitemsep,nolistsep]
\item Work available on Smartphones 
\item Worker selection possible by latitude/longitude 
\item Work matched to detailed worker profiles 
\item Verifiable worker identities 
\item QC by initial worker assessment and peer-review 
\item Cross-lingual website and tasks to attract German workers 
\end{itemize}
} & 

{ 
\begin{itemize} [noitemsep,nolistsep]
\item Rigorous worker screening and team work structure
\item QC detecting worker fatigue
\item Specialized support for transcription (via Humanoid and SpeakerText acquisitions
\item Pre-built task algorithms for use in hybrid workflows
\item Focus on ethical/sustainable practices
\item ML methods perform tasks and check human outputs with robot workers plugged into a virtual assembly line
\end{itemize}
} &

{
\begin{itemize} [noitemsep,nolistsep]
\item Machine automated workers
\item Automated quality \& cost optimization
\item Worker pool consisting of named and anonymous workers
\item Support for complex tasks, private crowds, and hybrid crowd
\item ML used to identify \& complete repetitive tasks, direct workers to the rest
\end{itemize}
} & 

{\begin{itemize} [noitemsep,nolistsep]
\item Broad demographics
\item Large workforce with task-targeting
\item Diverse incentives
\item Gold based QC
\end{itemize}
} & 

{\begin{itemize} [noitemsep,nolistsep]
\item Exclusive focus on writing tasks (write.com)
\item Supports organizational advancement
\end{itemize}
} &

{\begin{itemize} [noitemsep,nolistsep]
\item Supports hierarchical organization
\item Support for SMS-based mobile crowd work 
\item Support for Apple iOS application \& game testing on smart phones, and tablets
\item Detection \& prevention of task starvation
\end{itemize}
} &

{\begin{itemize} [noitemsep,nolistsep]
\item Support for complex tasks
\item Flexible, negotiated pay model (hourly vs. project-based)
\item Work-in-progress screenshots, time-sheets, and daily log
\item Public worker profiles (includes qualifications, work histories, past client feedback, test scores)
\item Rich communication \& mentoring support
\item Team room
\item Payroll/Health benefits support
\end{itemize}
}\\ 

\hline

{
\bf Whose Crowd?
} & 

{\begin{itemize} [noitemsep,nolistsep]
\item Own workforce
\end{itemize}
} &

{
\begin{itemize} [noitemsep,nolistsep]
\item Own workforce (discontinued use of AMT)
\item Supports private crowd as an Enterprise solution
\end{itemize}
} &

{
\begin{itemize} [noitemsep,nolistsep]
\item Workers come from AMT, eLance and oDesk
\item "CrowdVirtualizer" allows private crowds, including a combination of internal employees, outside specialists and crowd workers
\item Regular crowd workforce, private crowds supported
\item Active creation and addition of new public worker pools
\end{itemize}
} &

{
\begin{itemize} [noitemsep,nolistsep]
\item Own workforce; drawn fom 100+ channel partners
\item Supports private crowd at Enterprise level
\item Requesters can communicate directly with "contributers" via in-platform messaging
\end{itemize}
} &

{
\begin{itemize} [noitemsep,nolistsep]
\item Workers come from AMT
\item Subset of AMT workforce work on Crowdsource tasks
\end{itemize}
} &

{
\begin{itemize} [noitemsep,nolistsep]
\item Own workforce 
\item Workers and managers can recruit others
\item Private/closed crowds- not supported
\end{itemize}
} &

{
\begin{itemize} [noitemsep,nolistsep]
\item Own workforce
\item Private crowd supported under BYOC -"Bring your own contractor"
\end{itemize}

} \\ 
\hline

{\bf Demographic \& Worker Identities
} & 
{
\begin{itemize} [noitemsep,nolistsep]
\item  Most of the workers come from Europe, the US, and Canada
\item  TOS prohibit requesting personal info
\item  Workers can reveal PII to obtain "Trusted Member" status \& priority on project tasks
\item  Workers join online; payment via PayPal, enabling global recruitment
\end{itemize}
} &
{
\begin{itemize} [noitemsep,nolistsep]
\item  Nepal- based with Kenya pilot
\item  Has long- term goal of growing to 1M workers across 10-12 developing countries by 2018
\end{itemize}
} &
{
\begin{itemize} [noitemsep,nolistsep]
\item  Multiple workforce providers
\item  Broad demographics with access to both anonymous and known identity workers
\end{itemize}
} &
{
\begin{itemize} [noitemsep,nolistsep]
\item  Broad demographics due to multiple workforce providers
\item  Country \& channel information available at job order, whereas State and City targeting supported at Enterprise level
\item  Demographics targeting (gender, age, mobile- usage) available at Enterprise level
\item  Identities known to platform only for workers who create an optional CrowdFlower account
\end{itemize}
} &
{
\begin{itemize} [noitemsep,nolistsep]
\item  Workers come from 68 countries
\item  90\% reside in the US, 65\% have bachelor's degree or higher
\item  Education: 25\% doctorate, 42\% bachelors, 17\% some college, 15\% high school
\item  Other statistics: 65\% female, 53\% married, 64\% no children
\end{itemize}
} &
{
\begin{itemize} [noitemsep,nolistsep]
\item  Previously, workforce was primarily drawn from developing nations, but focus is no longer on the developing world
\item  Demographic information: NA
\item  Payments via PayPal, MoneyBookers or oDesk
\end{itemize}
} &
{
\begin{itemize} [noitemsep,nolistsep]
\item  Global workforce
\item  Public profile includes name, picture, location, skills, education, past jobs, tests taken, hourly pay rates, feedback, and ratings
\item  Workers' and requesters's identites are verified
\item  Variety of payment methods supported
\end{itemize}
} \\ 
\hline

{\bf Qualifications \&  Reputation
} & 
{
\begin{itemize} [noitemsep,nolistsep]
\item  Base and Project asessment test taken before beginning work
\item  Native \& Foreign language skills, expertise, and hobbies/interests data stored in profiles
\item  Workers' "Assessment Score" results are used to measure performance
\item  Continuous evaluation of skills based on work results
\end{itemize}
} &
{
\begin{itemize} [noitemsep,nolistsep]
\item  Workers hired after: 1) 30- min test via Facebook app; 2) Skype interview; 3) final online test
\item  Workers are assigned work after forming a team of 8 or joining an existing team
\item  Workers must take tests to qualify for work on certain tasks
\item  Further practice tasks are required until sufficient accuracy is established
\item  Periodic testing updates the accuracy scores
\item  Evaluation based on individual \& team performance
\end{itemize}
} &
{
\begin{itemize} [noitemsep,nolistsep]
\item  Performance compared to peers
\item  Workers are evaluated on gold tests, assigning each worker a score
\item  QC methods can predict worker accuracy
\end{itemize}
} &
{
\begin{itemize} [noitemsep,nolistsep]
\item  Skill restrictions can be applied
\item  Writing test, Wordsmith badge, indicates content writing proficiency
\item  Other badges shows specific skills, general trust resulting from work accuracy \& volume
\item  Skills/Reputation tracked via gold accuracy, number of jobs completed, type and variety of jobs completed, account age, and scores of skills tests
\item  Workers' broad interest \& expertise not recorded
\end{itemize}
} &
{
\begin{itemize} [noitemsep,nolistsep]
\item  Workers hired after a writing test, and undergo tests to qualify for tasks
\item  Reputation measured in terms of approval ratings, rankings, and tiers based on earnings
\end{itemize}
} &
{
\begin{itemize} [noitemsep,nolistsep]
\item  Workers' native language(s) recorded
\item  Workers complete certifications to gain access to restricted tasks
\item  Requesters can select required skills from a list while posting tasks
\item  Accuracy score, ranking \& number of completed tasks used to measure worker efficiency
\end{itemize}
} &
{
\begin{itemize} [noitemsep,nolistsep]
\item  Virtual interviews/chats with workers
\item  Work histories, past client feedback, test scores, and portfolios show worker qualifications and capabilities
\item  Tests can be taken to build credibility
\item  Worker- profiles include self reported English proficiency, and 10 areas of interest
\end{itemize}
} \\ 
\hline

{\bf Task Assignment \& Recommendations
} & 
{
\begin{itemize} [noitemsep,nolistsep]
\item  Each worker's task list is unique and shows tasks available based on Requester restrictions and worker profile compatibility
\end{itemize}
} &
{

\begin{itemize} [noitemsep,nolistsep]
\item  Tasks are individually dispatched based on prior performance
\end{itemize}
} &
{
\begin{itemize} [noitemsep,nolistsep]
\item  Automated methods can a) manage tasks; b) assign tasks to automated workers; c) manage workflows; d) worker contributitions
\end{itemize}
} &
{
\begin{itemize} [noitemsep,nolistsep]
\item NA
\end{itemize}
} &
{
\begin{itemize} [noitemsep,nolistsep]
\item NA
\end{itemize}
} &
{
\begin{itemize} [noitemsep,nolistsep]
\item  Dynamic task routing based on worker certifications \& accuracy scores, as well as task priority, though workers can select amongst these tasks

\item  Task priority used to reduce latency and prevent starvation
\end{itemize}
} &
{
\begin{itemize} [noitemsep,nolistsep]
\item  Jobs can be posted via a) Public post; b) Private invite; c) Both
\end{itemize}
} \\ 
\hline

{\bf Hierarchy \& Collaboration
} & 
{
\begin{itemize} [noitemsep,nolistsep]
\item   NA
\end{itemize}
} &
{
\begin{itemize} [noitemsep,nolistsep]
\item  Everyone works in a team of 5

\item  "Cloud Seeders" train new team leaders and each oversee 10-15 teams
\item  Team leaders lead weekly meetings and provide oversight, as well as character and leadership training
\item  Weekly meetings support knowledge exchange, task training, and accountability
\end{itemize}
} &
{
\begin{itemize} [noitemsep,nolistsep]
\item NA
\end{itemize}
} &
{
\begin{itemize} [noitemsep,nolistsep]
\item NA
\end{itemize}
} &
{
\begin{itemize} [noitemsep,nolistsep]
\item  Through tiered system, "virtual career system",  qualified workers gets promoted to become editors, who can be futher promoted to train other editors
\item  Community discussion forums let workers access resources, post queries, and contact moderators
\end{itemize}
} &
{
\begin{itemize} [noitemsep,nolistsep]
\item  Best workers can get promoted to managerial positions
\item  Experts check the work of other potential experts in a peer review system
\item  Experts also recruit new workers, evaluate potential problems with requester-defined tasks, and resolve disagreements
\item  Provides tools to enable worker-to-worker and manager-to-worker communication, including a worker chat interface
\end{itemize}
} &
{
\begin{itemize} [noitemsep,nolistsep]
\item  Rich communication supported between worker- requesters via the message center and team application
\item  Requester can build a team by hiring individual workers for the same project, and can optionally assign team management tasks to one of them
\item  Teams can be monitored using their "Team Room" application
\item  Requesters can view workers' latest screenshots, memos, and activity meters
\end{itemize}
} \\ 
\hline

\end{tabular}
\label{Table1}
\end{sidewaystable*}


\begin{sidewaystable*}
\tiny
    \centering
    \caption{Assessment of Platforms (continued)}
\begin{tabular}[h]{| p{1.5cm}| p{2.7cm}| p{2.7cm}| p{2.7cm}| p{2.7cm}| p{2.7cm}| p{2.7cm}| p{2.7cm}|}
\hline
{\bf Features} & {\bf Clickworker} &{\bf CloudFactory} & {\bf CrowdComputing Systems} &{\bf CrowdFlower} &{\bf CrowdSource} &{\bf MobileWorks} &{\bf oDesk} \\ 
\hline

{\bf Incentive Mechanisms
} & 
{
\begin{itemize} [noitemsep,nolistsep]
\item  \$5 referral bonus once newly joined worker earns \$10
\end{itemize}
} &
{
\begin{itemize} [noitemsep,nolistsep]
\item  Per-task payment

\item  Professional advancement, institutional mission, and economic mobility
\end{itemize}
} &
{
\begin{itemize} [noitemsep,nolistsep]
\item  Workers are paid relative to number of minutes spent, or number of tasks completed
\end{itemize}
} &
{
\begin{itemize} [noitemsep,nolistsep]
\item  Besides pay, badges \& optional leaderboard, use prestige and gamification

\item  Badges can give access to higher paying tasks and rewards
\item  Higher accuracy is rewarded through bonus payments
\end{itemize}
} &
{
\begin{itemize} [noitemsep,nolistsep]
\item  Virtual career system rewards top performers with higher pay, bonuses, awards and more access to work
\item  Rankings, in terms of total earnings \& total number of HITs, can be viewed for current month, year or total length of time on profile pages
\item  Graphical display of performance by different time periods is also available
\end{itemize}
} &
{
\begin{itemize} [noitemsep,nolistsep]
\item  Automatic setting of price/task based on estimated completion time, required effort, and worker hourly wages in local time zone ensuring fair pay
\item  Tiered payout based on worker accuracy
\end{itemize}
} &
{
\begin{itemize} [noitemsep,nolistsep]
\item  Workers set the hourly rate or work at fixed rate for a project
\item  Higher rated workers generally get more pay.
\item  For hourly pay model, Requesters pay only for the hours recorded in the "Work diary"
\end{itemize}
} \\ 
\hline

{\bf Quality Assurance \& Control
} & 
{
\begin{itemize} [noitemsep,nolistsep]
\item  QA: Specialized tasks, \& enterprise service, along with tests, continuous evaluation, job allocation according to skills and peer review
\item  Optional QC: internal QC checks, use of second- pass worker to proofread \& correct first- pass work products
\item  QC Methods: random spot testing by platform personnel, gold checks, worker agreement checks, plagiarism check \& proofreading
\item  Requesters \& Workers cannot communicate directly
\item  Rejected work needs to be redone by the worker and is checked for quality by platform personnel
\end{itemize}
} &
{
\begin{itemize} [noitemsep,nolistsep]
\item  Via Worker training, testing, and automated task algorithms
\item  Worker performance is monitored via the reputation systems
\item  Enterprise solutions develop custom online training that uses screen casts/shots to show corner cases and answer FAQs
\item  Practice tasks cover a variety of cases for training
\item  Workers are periodically assigned gold tests to assess their skills and accuracy
\end{itemize}
} &
{
\begin{itemize} [noitemsep,nolistsep]
\item  QA: Specialized tasks, enterprise- level offerings \& API support
\item  QC measures: monitoring keystrokes, time taken to complete the tasks, gold data, and assigning score
\item  Supports automated quality and cost optimization
\end{itemize}
} &
{
\begin{itemize} [noitemsep,nolistsep]
\item  QA: Rich communication via in- platform messaging, specialized tasks, \& enterprise- level offerings
\item  QC: Gold tests and skills/reputation management
\item  Tasks can be run without gold and platform will suggest gold for future jobs based on worker agreement
\item  Immediate feedback provided to workers on failing gold tests
\item  Requesters may suspend the job and re-assess gold if many workers fail gold tests
\item  "Hidden Gold" supported to minimize fraud via gold- memorization or collusion
\end{itemize}
} &
{
\begin{itemize} [noitemsep,nolistsep]
\item  Content review of writing tasks by editors (style \& grammar check)
\item  QC: Dedicated internal moderation teams check via plurality, algorithmic scoring, and gold checks
\item  Plagiarism check using CopyScape
\end{itemize}
} &
{
\begin{itemize} [noitemsep,nolistsep]
\item  Manual + Algorithmic techniques to manage quality including: dynamic work routing, peer management, social interaction techniques between manager-worker and worker-worker
\item  Worker reassigned to training tasks if worker accuracy, based on majority agreement, falls below the threshold
\item  Workers can request manager review in case of disagreement with majority answer
\item  Managers determine final answers for difficult examples
\end{itemize}
} &
{
\begin{itemize} [noitemsep,nolistsep]
\item  QA: "Work Diaries" report credentials, screenshots, screenshot metadata, webcam pictures, and number of keyboard \& mouse events
\item  QA: Rich communication supported via message center and team application
\item  No in-platform QC supported
\item  Enterprise solution provides QA through testing, certifications, training, work history and feedback ratings
\end{itemize}
} \\ 
\hline

{\bf Self-service, Enterprise, and API offerings
} & 
{
\begin{itemize} [noitemsep,nolistsep]
\item  Both, self- service and enterprise
\item  API support \& documentation available
\end{itemize}
} &
{
\begin{itemize} [noitemsep,nolistsep]
\item  Focus on enterprise solutions
\item  Earlier public API is now in private beta
\item  SpeakerText, acquired by CloudFactory, offers public API for transcription
\end{itemize}
} &
{
\begin{itemize} [noitemsep,nolistsep]
\item  Enterprise only, including an API
\item  Management tools include GUI, workflow management, and quality control
\end{itemize}
} &
{
\begin{itemize} [noitemsep,nolistsep]
\item  Both, self- service and enterprise
\item  "Basic" self- serve jobs can be custom built using GUI, while more technical jobs can be built using CML, CSS and Javascript
\item  "Pro" self-serve offering supports more quality tools, more skill groups and more channels
\item  Approximately 70 channels available for self- service, whereas all are available to "Pro" and enterprise jobs
\item  API support \& documentation available
\end{itemize}
} &
{
\begin{itemize} [noitemsep,nolistsep]
\item  Enterprise only
\item  Offer XML and JSON APIs
\end{itemize}
} &
{
\begin{itemize} [noitemsep,nolistsep]
\item  Enterprise only; self- serve option discontinued
\item  Self-service is replaced by virtual assistant service, "Premier", that supports small projects through post-by-email, expert finding system, and accuracy guarantee
\item  According to MobileWorks, allowing users to design tasks, communicate with workers on short-duration tasks is the wrong approach
\end{itemize}
} &
{
\begin{itemize} [noitemsep,nolistsep]
\item  Both, self-service and enterprise
\item  Numerous APIs supported thay may be used for microtasks as well
\end{itemize}
} \\ 
\hline

{\bf Specialized \& Complex Task Support
} & 
{
\begin{itemize} [noitemsep,nolistsep]
\item  Translation, web research, data categorization \& tagging, generating advertising text, SEO web content, and surveys
\item  Self-Service options provided for 2 tasks: generating advertising text and SEO web content

\item  Smartphone app
\end{itemize}
} &
{
\begin{itemize} [noitemsep,nolistsep]
\item  Transcription, data entry, processing, collection, and categorization
\end{itemize}
} &
{
\begin{itemize} [noitemsep,nolistsep]
\item  Modularized workflows
\item  Data structuring, content creation, content moderation, entity updation, search relevance improvement, and meta tagging
\item  Complex workflows decomposed into simple tasks, and repetitive tasks assigned to automated workers
\end{itemize}
} &
{
\begin{itemize} [noitemsep,nolistsep]
\item  Self-service API supports general and 2 specialized tasks: content moderation and sentiment analysis
\item  Enterprise level specialized tasks supported: business data collection, search result evaluation, content generation, customer and lead enhancement, categorization, and surveys
\item  Complex tasks supported at enterprise-level
\end{itemize}
} &
{
\begin{itemize} [noitemsep,nolistsep]
\item  Copywriting services, content tagging, data categorization, search relevance, content moderation, attribute identification, product matching, and sentiment analysis
\end{itemize}
} &
{
\begin{itemize} [noitemsep,nolistsep]
\item  Digitization, categorization, research, feedback, tagging, and others
\item  API support for workflows including parallel, iterative, survey and manual
\item  Tasks supported by API: natura-language response processing; image, text, language, speech and documents processing; dataset creation \& organization; testing; labeling; dataset classification
\end{itemize}
} &
{
\begin{itemize} [noitemsep,nolistsep]
\item  Arbitrarily complex tasks supported
\item  Web development, software development, networking \& information systems, writing \& translation, administrative support, design \& multimedia, customer service, sales \& marketing, business services
\item  Enterprise support for: writing, data entry \& research, content moderation, translation \& localization, software development, customer service, and custom solutions
\end{itemize}
} \\ 
\hline

{\bf Ethics \& Sustainability
} & 
{
\begin{itemize} [noitemsep,nolistsep]
\item  NA
\end{itemize}
} &
{
\begin{itemize} [noitemsep,nolistsep]
\item  Part of corporate mission
\item  Train each worker on character and leadership principles to fight poverty
\item  Teams assigned community challenges to serve others in need
\end{itemize}
} &
{
\begin{itemize} [noitemsep,nolistsep]
\item NA
\end{itemize}
} &
{
\begin{itemize} [noitemsep,nolistsep]
\item  Worker satisfaction studies occur regularly and have reportedly shown consistent high satisfaction
\end{itemize}
} &
{
\begin{itemize} [noitemsep,nolistsep]
\item  Seek to "provide fair pay, and competitive motivation to the workers"
\end{itemize}
} &
{
\begin{itemize} [noitemsep,nolistsep]
\item  Social mission is to employ marginalized population of developing nations
\item  Pricing ensures to provide fair or above- market hourly wages
\item  Workers can be promoted to become managers with more meaningful \& challenging work, hence supporting economic mobility
\end{itemize}
} &
{
\begin{itemize} [noitemsep,nolistsep]
\item  US and Canada based workers, working at least 30 hours/week are offered benefits including: simplified taxes, group health insurance, 401(k) retirement savings plan and unemployment benefits
\item  Requesters can benefit from platform's alternative to staffing firms and IRS compliance
\end{itemize}
} \\

\hline
\end{tabular}
\label{Table2}
\end{sidewaystable*}


{\bf Automated Task Algorithms.} What if any automated algorithms are provided to 
complement/supplement human workers~\cite{hu2011monotrans2}? 
Can Requesters inject their own automated algorithms into a workflow, blending human and automated processing? Are solutions completely automated for some tasks, or do they simplify the crowd's work by producing candidate answers which the crowd need only verify or correct? Some instances may be solved automatically, with more difficult cases routed to human workers. Work may be performed by both and then aggregated, or workers' outputs may be automatically validated to inform any 
subsequent review.
 
{\bf Ethics \& Sustainability.} What practices are adopted to promote an ethical and sustainable environment for crowd 
work~\cite{fort2011amazon,Irani13}? 
How are these practices implemented, assessed, and communicated to workers and Requesters? How does the platform balance ethics and sustainability against competing concerns in a competitive market where practical costs tend to dominate discussion and drive adoption?


\section{Assessment of Platforms}

The previous Section defined the categories we developed to characterize and differentiate crowd work platforms for our content analysis. We now present our deductive application of these categories to seven crowd work platforms: ClickWorker, CloudFactory, CrowdFlower, CrowdComputing Systems, CrowdSource, MobileWorks, and oDesk. To avoid overly simplistic and polemic comparisons, we focus on assessing platform capabilities with respect to each category. Our assessment appears in Tables~\ref{Table1} and \ref{Table2}. 

After reviewing available commercial platforms (cf.~\cite{ipeirotis-micro-services,ipeirotis-verticals}), the platforms above were selected based on a combination of factors: connections to the research community, popularity of use, resemblance to AMT's general model and workflow while still providing significant departures from it, and collectively encompassing a range of diverse attributes. In comparison to the six platforms in Turian's analysis~\cite{Turian12}, we exclude Servio (\url{www.serv.io}) and add ClickWorker and oDesk. As an online contractor marketplace, oDesk both provides contrast with microtask platforms and reflects prior interest and work in the research community~\cite{Ipeirotis-odesk}. We exclude AMT here, assuming readers are already familiar with it, though we provide contrasts in the next section. While other industrial white papers have covered many more platforms (cf.~\cite{frei2009paid}), the level of analysis presented in such cases tends to be quite shallow, with little clarity as to what criteria mean, let alone how they were assessed. We focus here on depth and quality of coverage to ensure our analysis provides meaningful understanding of platform capabilities and how we assessed them.

Our content analysis assesses for each platform by drawing upon a variety of information sources (Webpages, blogs, news articles, white papers, and research papers). We also shared our analysis of each platform with that platform's representatives and requested their feedback. Four of the seven platforms provided feedback, which we incorporate and cite. In case of insufficient information to assess a given platform-category combination, we enter "NA". We use the following acronyms in the table: ML (Machine Learning), QA (Quality Assurance), QC (Quality Control), TOS (Terms of Service), CML (CrowdFlower Mark-up Language). For the interested reader, further details for each platform are in the Appendix.  


\section{Discussion}

AMT helped give rise to crowd work research and continues to shape research questions and directions being investigated. In prior work, researchers have reported many challenges with AMT that have made use more difficult and often led to research seeking to address current limitations. 


For example, we earlier cited many studies that have sought to improve complexity and quality of crowd work performed on AMT, but what AMT assumptions underlie such work? Crowd workers are often treated as interchangeable and unreliable. Specialized skills or knowledge possessed by workers is often ignored, high-rates of attrition are expected, and only simple tasks requiring short attention spans are offered. By assuming workers are interchangeable, we lose the opportunity to benefit from their diverse skills and knowledge. When workers are not trusted, we must continually verify the quality of their work. When we assume workers are capable of only simple tasks, we must carefully decompose complex tasks or abandon them. When we assume workers may come and go at any time, we cannot perform tasks requiring greater continuity. When untrusted workers provide unexpected responses, we may question the quality of their work rather than benefit from their diversity of opinion. How might these studies' approaches and findings have differed if they had instead assumed a platform where identities and skill profiles were known, or where work was matched to workers, or where workers were paid hourly rather than piecemeal and were offered opportunities for advancement, or where complex tasks were directly supported? The possibilities are vast.

The previous section reviewed and assessed seven alternative crowd work platforms, highlighting their distinguishing features. In this section, we discuss these platform capabilities to identify new insights or opportunities they provide which suggest a broader perspective on prior work, revised methodology for effective use, and directions for future research.  

{\bf Whose Crowd?} As discussed by Turian~\cite{Turian12}, some platforms believe curating their own workforce is key to QA, while others rely on other workforce-agnostic QA/QC methods. Like AMT, platforms such as ClickWorker, CloudFactory, CrowdSource, MobileWorks, and oDesk have their own workforce. However, unlike AMT, where anyone can join, these platforms (except oDesk) vet their workforce through screening or by only letting workers work on tasks matching their backgrounds/skills. 
On the other hand, for cases in which using ones's own private crowd is necessary or desired, several platforms offer enterprise-level support: 
CloudFactory, CrowdComputing Systems, CrowdFlower, and oDesk.

{\bf Demographics \& Worker Identities.} AMT's workforce is concentrated in the U.S.\ and India, with limited demographic filtering and lack of worker identities. If AMT no longer accepts new international workers~\cite{no-intl}, over time this could adversely impact its demographics, scalability, and latency.

All the platforms we considered provide access to an international workforce, with some targeting specific regions. ClickWorker has most of the workers coming from Europe, US, and Canada, while Nepal-based CloudFactory draws its workforce from developing nations like Nepal and Kenya, and MobileWorks focuses on India, Jamaica, and Pakistan. While CrowdComputing draws its workforce from AMT, eLance and oDesk, CrowdFlower may have the broadest workforce of all through partnering with many workforce providers. On the other hand, for those needing US-only workers, 90\% of CrowdSource's workers hail from the U.S. 

Some tasks necessitate selecting workers with specific demographic attributes, e.g.\ usability testing (\url{www.utest.com}). Several platforms offer geographic restrictions for focusing tasks on particular regions, e.g., CrowdFlower supports targetting by city or state, while ClickWorker allows latitude and longitude based restrictions. While further demographic restrictions may be possible for conducting surveys, this is rarely available across self-service solutions, perhaps due to reluctance to provide detailed workforce demographics. This suggests the need to collect demographic information external to the platform itself will likely continue~\cite{Ross10,ipeirotis2010demographics}.  

Whereas AMT lacks worker profile pages where workers' identity and skills can be showcased, creating a ``market for lemons''~\cite{Ipeirotis-lemons} and leadings some researchers to pursue social network integration~\cite{difallah13},
oDesk offers public worker profiles displaying their identities, skills, feedback, and ratings information. ClickWorker has a pool of ``Trusted members" with verified IDs along with anonymous members. However, for tasks such as human subjects research \cite{mason2012conducting}, knowledge of worker identities may actually be undesirable. What appears lacking presently is any platform offering both options: programmable access to pull identities on-demand for tasks that require greater credibility, yet the ability to hide these identities when they are not desired. 




{\bf Qualifications \& Reputation.} As discussed earlier, limitations of AMT's reputation system are well known~\cite{panos2010pleatoamazon}. Requesters design their own custom qualification tests, depend on semi-reliable approval ratings~\cite{ipeirotis-be-a-top-worker}, or use pre-qualified ``Master'' workers. Because their is no common ``library'' of tests~\cite{chen2011opportunities}, each requester must write their own, even for a similarly defined tasks. No pre-defined common tests to check frequently tested skills or to measure language proficiency~\cite{mellebeek2010opinion} exist on AMT. On the other hand, custom qualification tests provide Requesters great control and flexibility in assessing workers for their particular custom tasks. 

In contrast, ClickWorker, CloudFactory, CrowdSource, and MobileWorks test workers before providing them relevant work. ClickWorker makes their workers undergo base and project assessment tests before beginning work. CrowdComputing Systems and CrowdFlower test workers via gold tests. Besides these gold tests, through CrowdFlower, Requesters can apply skill restrictions on tasks which can be cleared by taking platform�s standardized skill tests. eg. writing, sentiment analysis, etc. CrowdSource creates a pool of pre-qualified workers by hiring workers only after they pass a writing test. Like CrowdFlower, MobileWorks awards certifications for content creation, digitization, internet research, software testing, etc. after taking lessons and passing the qualification tests to earn access to restricted tasks.
CloudFactory and oDesk allow more traditional screening practices in hiring workers. In CloudFactory, workers are hired only on clearing tests taken via Facebook app, Skype interview, and an online test. Whereas, oDesk that follows contract model, allows requesters to select workers through virtual interviews, and test scores on platform defined tests such as US English basic skills test, office skills test, email etiquette certification, call center skills test, search engine optimization test, etc. 


Without distinguishing traits, workers on AMT appear interchangeable and lack differentiation based on varying ability or competence~\cite{Ipeirotis-lemons}. 
CrowdFlower and MobileWorks uses badges to display workers' skills. CrowdSource, and CrowdFlower additionally use a leaderboard to rank workers. Worker profiles on oDesk display work histories, feedback, test scores, ratings, and areas of interests that helps enable requesters to choose workers matching their selection criteria.

Lack of worker analytics on AMT has also led to a variety of research~\cite{heymann2011turkalytics,Rzeszotarski11}. While CrowdFlower and others are increasingly providing more detailed statistics on worker performance, this continues to be an area in need of more work.

{\bf Task Assignments \& Recommendations}. On AMT, workers can only search for tasks by keywords, payment rate, duration, etc., making it more difficult to find tasks of interest~\cite{chilton2010task,law2011effects,panos2010pleatoamazon}. Moreover, AMT does not provide any support to route tasks to appropriate workers \cite{Ho2012}. This has prompted some researchers to try to improve upon AMT's status quo by designing better task search or routing mechanisms. 

However, ClickWorker already shows a worker only those tasks for which he is eligible and has passed the relevant qualification test. Since a worker is given an option to take qualification tests per his interests, there is a high probability that the subsequent tasks are interesting to him. MobileWorks goes further by routing tasks algorithmically based on workers accuracy scores and certifications. It also maintains priority of the tasks as well to reduce task starvation. CrowdComputing Systems similarly uses algorithmic task routing. On oDesk, Requesters can post tasks as public, private, or hybrid. 
  
{\bf Hierarchy \& Collaboration}. On oft-discussed limitation of AMT is lack of native support for traditional hierarchical management structures~\cite{kochhar2010anatomy,Nallapati13}. AMT also lacks worker collaboration tools, other than online forums. 
Interaction may enable more effective workers to manage, teach, and assist other workers. This may help the crowd to collaboratively learn to better solve new tasks~\cite{Kulkarni12}. 

ClickWorker, CrowdSource, MobileWorks, and CloudFactory support peer review. MobileWorks and CloudFactory promote workers to leadership positions. 
CrowdSource also promotes expert writers to editor and trainer positions. 
With regard to collaboration, 
worker chat (MobileWorks, oDesk), forums (CrowdFlower, ClickWorker, CrowdSource, oDesk) and Facebook pages support worker-requester and worker-platform interactions. Whereas some researchers have developed their own crowd systems to support hierarchy~\cite{kochhar2010anatomy,Nallapati13} or augmented AMT to better support it, we might instead exploit other platforms where collaboration and hierarchy are assumed and structures are already in place. This point may seem obvious, yet we still see new studies which describe AMT's lack of support as the status quo to be rectified. 





{\bf Incentive Mechanisms}. AMT's incentive architecture lets Requesters specify piecemeal payments and bonuses, with little guidance offered in pricing tasks for effectiveness or fairness~\cite{Mason:Watts:2009,faridani2011s,singer2011pricing}. 
How often might limitations of this incentive model underly poor quality or latency researchers have reported or sought to address in prior work?  What if they had simply assumed an alternative platform's model instead?

Standard platform-level incentive management can help ensure that every worker doing a good job is appropriately rewarded. 
CrowdFlower pays bonuses to workers with higher accuracy scores. CrowdSource has devised a virtual career system that pays higher wages, bonuses, awards, and access to more work to deserving workers. MobileWorks follows tiered payment method where workers whose accuracy is below 80\%  earns only 75\% of the overall possible earnings. oDesk, like Metaweb \cite{kochhar2010anatomy}, allows hourly wages to workers, giving workers the flexibility to choose their own hourly rate according to their skills and experience.

While AMT focuses exclusively on payment incentives at the platform level, CrowdFlower and MobileWorks now provide badges which recognize workers' achievements. CrowdSource, and CrowdFlower additionally provide a leaderboard for the workers to gauge themselves against peers. 
Relatively little research to date or existing platforms have explored generalizable mechanisms for effectively integrating other gamification mechanisms with crowd work. Some platforms motivate quaity work through opportunities for skill acquisition and professional and economic advancement, as discussed under the other categories here. 


{\bf Quality Assurance \& Control}. AMT provides only minimal QA and QC 
\cite{Ipeirotis:Wang:2010,dow2012shepherding}. For QC, requesters have to insert trap questions or devise anti-robot tests~\cite{negri2010creating,zhu10,sanderson2010user}. Open questions regarding requester trustworthiness and fraud remain unanswered with AMT \cite{panos2010pleatoamazon,tn-avoid}. Worker fraud and use of robots on AMT disadvantages all other parties~\cite{levine2006survey,Kittur08,downs2010your,heymann2011turkalytics,difallah2012mechanical,raykar2012eliminating,Eickhoff13}. While many statistical QC algorithms have been published, there has been only minimal discussion of how various underlying AMT assumptions may be a significant source of quality problems.

Clickworker uses peer review, plagiarism check, and testing. MobileWorks uses dynamic work routing, peer management, and social interaction techniques, with native workflow support for QA. oDesk uses testing, certifications, training, work history and feedback ratings. Other platforms, such as CrowdFlower and CrowdSource, focus on integrating and providing standardized QC methods, rather than placing the burden on Requesters. CrowdFlower makes native use of gold tests to filter out low quality results and spammers. CrowdSource uses plurality, algorithmic scoring, plagiarism check and gold checks. CrowdComputing Systems monitors keystrokes, time taken to complete the task, gold data, and assigns score as part of their QC checks. 

Our field would benefit from not tackling QA/QC from scratch and/or comparing to an AMT baseline. Instead, future work should utilize and compare to the foundations offered by alternative platforms, providing much needed diversity in comparison to prior AMT QA/QC research.





{\bf Self-service, Enterprise, and API offerings}. AMT supports both self-service and enterprise solutions along with access to their API. All other platforms 
in our study offer API support with different levels of customization. Besides facilitating task design and development,  platforms offer RESTful APIs supporting features such as custom workflows, data formats, environments, worker selection parameters, development platforms, etc. Lack of a better interface to post tasks \cite{panos2010pleatoamazon} has led requesters to use toolkits ~\cite{little2009turkit,kittur2011crowdforge}. Future research could devise and qualitatively compare alternative crowd programming models for performing a benchmark of different tasks, assessing where current APIs across platforms are sufficient, innovative differences between APIs, and where better API architectures and programming models could ease or enhance work vs.\ current offerings.  The Appendix provides further detail on API offerings. 


{\bf Specialized \& Complex Task Support}. Prior studies have often pursued complex strategies with AMT in order to produce quality output, e.g., for common tasks such as transcription~\cite{novotney2010cheap} or usability testing~\cite{Liu12-asist}. However, other platforms 
provide better tools to design a task, or pre-specified tasks with a workforce trained for them (e.g., CastingWords for transcription, uTest for usability testing, etc.). ClickWorker provides support for a variety of tasks but with special focus on SEO text building tasks. CrowdSource supports writing \& text creation tasks, and also provides resources to workers for further training. MobileWorks enables testing of iOS apps and games. They also support mobile crowdsourcing by making available tasks that can be completed via SMS. 
Using oDesk, requesters can get complex tasks done such as web development, software development, etc. Enhanced collaboration tools available on some platforms enable workers to better tackle complex tasks.


{\bf Automated Task Algorithms}. While machine learning methods could be used to perform certain tasks and verify results, AMT does not provide or allow machine learning techniques to be used for task performance. Only human workers are supposed to perform the tasks, irrespective of how repetitive or monotonous they may be. In contrast, CrowdComputing Systems allows usage of  machine automated workers,  identifies pieces of work that can be handled by automated algorithms, and uses human judgments for the rest. This enables better task handling and faster completion times. CloudFactory allows using robot workers to plug into a virtual assembly line. 
More research is needed in such ``hybrid'' crowdsourcing to navigate the balance and effective workflows for integrating machine learning methods and human crowds together for optimal task performance.

{\bf Ethics \& Sustainability}. AMT has been called a ``digital sweatshop''~\cite{Cushing13}, and some in the research community have raised ethical concerns regarding our implicit support of 
common AMT practices~\cite{Silberman10,fort2011amazon,Irani13,adda2013economic}. Mason and Suri discuss researchers' uncertainty in how to price tasks fairly in an international market~\cite{mason2012conducting}. 
However, unlike AMT, some other crowdsourcing platforms now promise humane working conditions and/or living wages: e.g., CloudFactory, MobileWorks~\cite{narula2011mobileworks,Kulkarni12}, and SamaSource. Focus on work ethics can be one of the motivating factors for the workers. oDesk, on the other hand, provides payroll and health-care benefits like the traditional organizations. It is unlikely that all work can entice volunteers or be gamified, and free work is not clearly more ethical than paying for it~\cite{fort2011amazon}. These other platforms offers ways to imagine a future of ethical, paid crowd work.

\section{Conclusion}

While AMT has had tremendous impact upon the crowdsourcing industry and research studies and practices, AMT remains in beta  with a number of well-known limitations. This represents both a bottleneck and risk of undue influence to ongoing research directions and methods. We suggest it would benefit research on crowd work to further diversify its focus to provide greater attention to alternative platforms as well.

In this paper, we developed a set of useful categories to characterize and differentiate alternative crowd work platforms. 
On the basis of these categories, we conducted a detailed study and assessment of seven crowdsourcing platforms: ClickWorker, CloudFactory, CrowdFlower, CrowdComputing Systems, CrowdSource, MobileWorks, and oDesk. By examining the range of capabilities and work models offered by different platforms, and providing contrasts with AMT, our analysis informs both methodology of use and directions for future research. Our cross-platform analysis represents the only such study we are aware of by researchers for researchers, intended to foster more study and use of AMT alternatives in order to further enrich research diversity. 
Our analysis identified some common practices across several platforms, such as peer review, qualification tests, leaderboards, etc. At the same time, we also saw various distinguishing approaches, such as use of automated methods, task availability on mobiles, ethical worker treatment, etc.


The scene of research and development in the crowd work industry is changing rapidly, and some open problems identified in prior research still remain unsupported by today's platforms. For instance, lack of support for {\em flash crowds}, i.e., a group of individuals who arrive moments after a request and can work synchronously~\cite{Kittur13}. Varying platform support for traditional organizational concepts of hierarchy and collaboration merits further analysis, particularly performing complex tasks effectively. 
Also, crowd work largely remains a desktop affair, with diminishing attention to mobile work~\cite{eagle2009txteagle,mw-kulkarni13}. 
Further work is also needed to manage identity sharing, e.g., to pull identities on-demand for tasks that need greater credibility, and yet be able to hide them when they are not desired. Similarly, the need to collect and study crowd demographics outside of platforms will likely continue~\cite{Ross10,ipeirotis2010demographics}. 

While several platforms offer multiple incentives to workers, it is not clear how to best utilize such incentives in combination, and prior research has focused primarily on study effects of isolated incentives rather than their use in combination. Across platforms we still see insufficient support for well-designed worker analytics~\cite{heymann2011turkalytics,Rzeszotarski11}. Regarding QA/QC, future research could benefit by not building from and/or comparing to an AMT baseline. Instead, we might utilize foundations offered by alternative platforms, providing valuable diversity in comparison to prior AMT QA/QC research. In regard to programming models, future research could usefully compare alternative APIs and programmatic workflows for performing a benchmark of different tasks, assessing where current APIs across platforms are sufficient, innovative differences between APIs, and where better API architectures and programming models might ease or enhance work vs.\ current offerings. More study of ``hybrid'' crowdsourcing could help us better design effective workflows for integrating automatic methods and human crowds for optimal task performance.
 
{\bf Acknowledgments.} This research was supported in part by DARPA Award N66001-12-1-4256 and IMLS grant RE-04-13-0042-13. Any opinions, findings, and conclusions or recommendations expressed by the authors do not express the views of any of the supporting funding agencies.

\balance

\bibliographystyle{acm-sigchi}
\bibliography{references}

\section{APPENDIX}

The earlier {\em Assessment of Platforms} Section presented our deductive application of analysis categories, concisely presenting the results of our assessment in Tables~\ref{Table1} and \ref{Table2}. In this (optional) appendix, we provide the interested reader significantly more detail regarding our assessment of each platform.

\subsection{ClickWorker} 

{\bf Distinguishing Features}. Provides work on smartphones; can select workers by latitude/longitude; work matched to detailed worker profiles; verify worker identities; control quality by initial worker assessment and peer-review; attract German workers with cross-lingual website and tasks.


{\bf Whose Crowd?} Provides workforce; workers join online. 

{\bf Demographics \& Worker Identities.} While anyone can join, most workers come from Europe, the US, and Canada. Terms-of-service (TOS) prohibit requesting personal information from workers.  
Tasks can be restricted to workers within a given radius or area via latitude and longitude. 
Workers can disclose their real identities (PII) for verification to obtain ``Trusted Member'' status and priority on project tasks.

{\bf Qualifications \& Reputation.} Workers undergo both base and project assessment before beginning work. Native and foreign language skills are stored in their profiles, along with both expertise and hobbies/interests. Workers' {\em Assessment Score} results are used to measure their performance. Skills are continuously evaluated based on work results.

{\bf Task Assignment \& Recommendations}. Each worker's unique task list shows only those tasks available based on Requester restrictions and worker profile compatibility (i.e., self-reported language skills, expertise, and interests). 

{\bf Hierarchy \& Collaboration.} 
Workers can 
discuss tasks and questions on platform's online forum
and Facebook group.

{\bf Incentive Mechanisms.} 
Workers  
receive \$5 US for each referred worker who joins and earns \$10 US. Workers are paid via PayPal, enabling global recruitment.

{\bf Quality Assurance \& Control} (QA/QC). API documentation, specialized tasks, and enterprise service promote quality, along with tests, continual evaluation, and job allocation according to skills and peer review. Optional QC includes  
use of internal QC checks or use of a second-pass worker to proofread and correct first-pass work products. QC methods include: random spot testing by platform personnel, gold checks, worker agreement checks (majority rule), plagiarism check, 
and proofreading. 
Requesters and workers cannot communicate directly. Work can be rejected with a detailed explanation, and only accepted work is charged. The worker will then be asked to do it over again. Quality is then checked by platform personnel, who decide if it is acceptable or if a different worker should be asked to perform the work. 




{\bf Self-service, Enterprise, and API offerings}. Self-service and enterprise are available. Self-service options are provided for generating advertising text and SEO web content. RESTful API offerings: customizable workflow; production, sandbox (testing), and sandbox beta (staging) environments supported; localization support; supports XML and JSON; no direct access to workers, however filters can be applied to select a subset of workers based on requested parameters, e.g., worker selection by longitude \& latitude. 
API libraries support Wordpress-SEO-text plugin, timezone, Ruby templates, Ruby client for single sign-on system, and Rails plugin. 


{\bf Specialized \& Complex Task Support}. Besides generating advertising text and SEO web content, other specialized tasks include: translation, web research, data categorization \& tagging, and surveys. 
A smartphone application enables geo-coding of photos and collection and verification of local data. 




\subsection{CloudFactory}

{\bf Distinguishing Features.} Rigorous worker screening and team work structure; QC detecting worker fatigue and specialized support for transcription (via Humanoid and SpeakerText acquisitions); pre-built task algorithms for use in hybrid workflows; focus on ethical/sustainable practices.

{\bf Whose Crowd?} Provides own workforce; they discontinued use of AMT over a year ago~\cite{cf-sears13}. 
They offer private crowd support as an Enterprise solution.

{\bf Demographics \& Worker Identities.} Nepal-based with Kenya pilot; plan to aggressively expand African workforce in 2013; longer-term goal is to grow to one million workers across 10-12 developing countries by 2018~\cite{Shu-cloudfactory}.

{\bf Qualifications \& Reputation.} Workers are hired after: 1) a 30-minute test via a Facebook app; 2) a Skype interview; and 3) a final online test~\cite{Knight13}. Workers are assigned
work after forming a team of 8 or joining an existing team. Workers must take tests to qualify for work on certain tasks. Further practice tasks are required until sufficient accuracy
is established
. Periodic testing updates these accuracy scores. Workers are evaluated by both individual and
team performance. 




{\bf Hierarchy \& Collaboration.} Tasks are individually dispatched 
based on prior performance; everyone works in teams of 5~\cite{cf-sears13}.
{\em Cloud Seeders} train new team leaders and each oversee 10-20 teams of 5 workers each
. Team leaders lead weekly meetings and provide oversight, as well as character and leadership training~\cite{Strait12,Sears-quora}. Weekly meetings support knowledge exchange, task training, and accountability. 

{\bf Incentive Mechanisms.} Per-task payment; professional advancement, institutional mission, and economic mobility. 

{\bf Quality Assurance \& Control.} Quality is managed via worker training, testing, and automated task algorithms. 
Worker performance is monitored via the reputation systems. 
Enterprise solutions 
develop custom online training that uses screen casts/shots to show corner
cases and answer frequently asked worker questions. 
Practice tasks cover a variety of cases for training. Workers are periodically assigned gold tests to assess their skills and accuracy. 


{\bf Self-service, Enterprise, and API offerings}.  They focus on Enterprise solutions. Earlier public API is now in private beta, though the acquired SpeakerText platform does offer a public API. 
API offerings: RESTful API; supports platforms such as YouTube, Vimeo, Blip.tv, Ooyala, SoundCloud, and Wistia; supports JSON; API library available.

{\bf Specialized \& Complex Task Support}. Specialized tasks supported include transcription and data entry, processing, collection, and categorization. 

{\bf Automated Task Algorithms}. Machine learning methods perform tasks and check human outputs. Robot workers
can be plugged into a virtual assembly line to integrate functionality, including Google Translate, media conversion and splitting, email generation, scraping content, extracting
entities or terms, inferring sentiment, applying image filters, etc. Acquired company Humanoid used machine learning to infer accuracy of work as a function of prior history and other factors, such as time of day and worker location/fatigue~\cite{Humanoid}.
%

{\bf Ethics \& Sustainability}. Part of mission. CloudFactory trains each worker on character and leadership principles 
to fight poverty. 
Teams are assigned a community challenge where they need to go out and serve others who are in need. 

\subsection{CrowdComputing Systems} 

{\bf Distinguishing Features.} Machine automated workers, automated quality \& cost optimization, worker pool consisting of named and anonymous workers, support for complex tasks, private crowds, and 
hybrid crowd via {\em CrowdVirtualizer}.

{\bf Whose Crowd?} 
Uses workers from AMT, eLance and oDesk. 
{\em CrowdVirtualizer} allows private crowds, including a combination of internal employees, outside specialists and crowd workers. Both regular crowd workforce and private crowds are used, plus automated task algorithms. They are actively creating and adding new public worker pools, e.g., 
CrowdVirtualizer's plug-in would let Facebook or CondeNast easily offer crowd work to their users~\cite{ccs-devine13}.

{\bf Demographics \& Worker Identities.} Using multiple workforce providers 
facilitates broad demographics with access to both anonymous and known identity workers. 

{\bf Qualifications \& Reputation.} 
Worker performance is compared to peers and evaluated on gold tests, assigning each worker a score. 
Methods below predict worker accuracy. 

{\bf Task Assignment \& Recommendations}. 
Automated methods manage tasks, assign tasks to automated workers, 
and manage workflows and worker contributions. 


{\bf Incentive Mechanisms.}  Crowd workers are paid for the tasks by verifying the effort after either checking the total number of minutes  each worker spent on a task, or by noting the number of tasks completed by the worker.  

{\bf Quality Assurance \& Control.} Specialized tasks, enterprise-level offerings, and API support QA. The quality control measures include monitoring keystrokes, time taken to complete the task, gold data, and assigning score. Supports automated quality and cost optimization~\cite{ccs-devine13}. 

{\bf Self-service, Enterprise, and API offerings}. Enterprise-only solutions include an API. 
Enterprise management tools 
include GUI, workflow management, and quality control.

{\bf Specialized \& Complex Task Support}. Workflows are modularized so that the modules can be reused in some other work processes. Enterprise offerings for specialized tasks include: structuring data, creating content, moderating content, updating entities, improving search relevance, and meta tagging. 
They decompose complex workflows into simple tasks and assign repetitive tasks to automated workers. 

{\bf Automated Task Algorithms}. 
Machine learning is used to identify and complete repetitive tasks, then direct workers to remaining tasks that require human judgment. For example, 
using machine workers 
for information extraction and human workers to find and interpret this information~\cite{CCSBlog}.  



\subsection{CrowdFlower}

{\bf Distinguishing Features}. Broad demographics, large workforce with task-targeting, diverse incentives, and gold-based QC~\cite{Le10,Oleson11}.

{\bf Whose Crowd?} 
Workforce drawn from 100+ channel partners~\cite{cf-josephy13} (public and private). 
Private crowds are supported 
at the enterprise-level. Requesters may interact directly with ``contributors'' via in-platform messaging.  

{\bf Demographics \& Worker Identities.} Channel partners provide broad demographics. Country and channel information is available at job order. State and city targeting are supported at the enterprise-level. Demographics targeting  (gender, age, mobile-usage) is also available at the enterprise-level. Worker identities are known (internal to the platform) for those workers who elect to create an optional CrowdFlower account that lets them track their personal performance across jobs~\cite{cf-josephy13}.


{\bf Qualifications \& Reputation.} 
Skill restrictions can be specified. 
A writing test {\em WordSmith} badge indicates 
content writing proficiency. 
Other badges recognize specific skills (e.g., Pic Patrol, Search Engineer, and Sentiment), general trust resulting from work accuracy and volume
 (Crowdsourcerer, Dedicated, and Sharpshooter), and experts who provide usability feedback on tasks as well as answers in the CrowdFlower support forum
 ({\em CrowdLab}).
Skills/reputation is tracked in a number of ways, including gold accuracy, number of jobs completed, type and variety of jobs completed, account age, and scores of skills tests created by requesters~\cite{cf-josephy13}. 
Workers' broad interests and expertise are not recorded.



{\bf Incentive Mechanisms.} Besides pay, badges and an optional leaderboard use prestige and gamification. 
{\em CrowdLab} badge recognition may promote altruism. 
Badges can give access to higher paying jobs and rewards [2]. Bonus payments called {\em CrowdCred} are offered for achieving high accuracy. 
Worker-satisfaction studies are reported to occur regularly and consistently show high worker satisfaction \cite{eastbayexpress}.

{\bf Quality Assurance \& Control.} QA is promoted via API documentation and support, specialized tasks, and enterprise-level offerings. QC is primarily based on gold tests and skills/reputation management. 
Requesters can run a job without gold and the platform will suggest gold for future jobs based on worker agreement. 
Workers are provided immediate feedback upon failing gold tests. 
{\em Hidden gold} can also be used, in which case workers are not informed of errors, making fraud via gold-memorization or collusion more difficult. 
If many workers fail gold tests, the job is suspended so the Requester may re-assess gold~\cite{finin2010annotating,negri2010creating}.

{\bf Self-service, Enterprise, and API offerings}. Basic self-service platform allows building custom jobs using a GUI. More technical users can build jobs using CML (CrowdFlower Mark-up Language) paired with custom CSS and Javascript. A ``Pro'' offering provides access to more quality tools, more skills groups, and more channels.  
The self-service API 
covers general tasks and two specialized tasks: content moderation 
and sentiment analysis. 
70 channels are available for self-service, while all are available to Platform Pro and enterprise jobs~\cite{cf-josephy13}. 
RESTful API offerings: 
{\em CrowdFlower API}: JSON responses; 
bulk upload (JSON), data feeds (RSS, Atom, XML, JSON), and spreadsheets; add/remove of gold-quality checking units; setting channels; 
{\em RTFM (Real Time Foto Moderator)}: Image moderation API; JSON.

{\bf Specialized \& Complex Task Support}. 
They also offer specialized enterprise-level solutions for business data collection, search result evaluation, content generation, customer and lead enhancement, categorization, and surveys. Support for complex tasks is addressed at the enterprise-level. 



\subsection{CrowdSource} 
{\bf Distinguishing Features.} Focused exclusively on vertical of writing tasks (\url{write.com}), supports economic mobility.

{\bf Whose Crowd?} AMT solution provider; a subset of AMT workforce is approved to work on CrowdSource tasks. 

{\bf Demographics \& Worker Identities.} Report 90\% of their workforce reside in the US and 65\% have bachelor's degrees or higher~\cite{cs-capabilities}. Workers come from 68 countries. 
Additional statistics include: education (25\% doctorate, 42\% bachelors, 17\% some college, 15\% high school); 65\% female, 53\% (all workers) married; 64\% (all workers) report no children.  

{\bf Qualifications \& Reputation.} Workers are hired after a writing test and  
undergo a series of tests to qualify for tasks. Reputation is measured in terms of approval rating, ranking, and tiers based on earnings (e.g., the top 500 earners). 


{\bf Hierarchy \& Collaboration.} 
A tiered system {\em virtual career system} lets qualified workers apply and get promoted to become editors, who can be further promoted to train other editors~\cite{CSScalableSol}. Community discussion forums
 let workers access resources, post queries, and contact moderators.

{\bf Incentive Mechanisms.} 
The {\em virtual career system} rewards top-performing workers with higher pay, bonuses, awards and access to more work. On worker profile pages, workers can view their rankings in two categories: total earnings and total number of HITs. Rankings can be viewed for current month, current year or total length of time. Graphs depicting performance by different time periods are also available~\cite{CSWorkerProfiles}.

{\bf Quality Assurance \& Control.} Editors review content writing tasks and ensure that work adheres to their style guide and grammar rules. Tasks are checked for quality by dedicated internal moderation teams via plurality, algorithmic scoring, and gold checks. 
Plagiarism is checked using CopyScape. 

{\bf Self-service, Enterprise, and API offerings}. They provide enterprise level support only. They offer XML and JSON APIs to integrate solutions with client systems.

{\bf Specialized \& Complex Task Support}. They provide support for copywriting services, content tagging, data categorization, search relevance, content moderation, attribute identification, product matching, and sentiment analysis. 


{\bf Ethics \& Sustainability}. They seek to ``provide fair pay, and competitive motivation to the workers''~\cite{CSadditionWP}. 

\subsection{MobileWorks}

{\bf Distinguishing Features.} Support for mobile crowd work via SMS optical character recognition (OCR) tasks on low-end phones (though only a small minority of their workforce still uses mobile~\cite{mw-kulkarni13}), as well as Apple iOS application and game testing on smart phones and tablets; support hierarchical organization; detect and prevent task starvation. 

{\bf Whose Crowd?} Have own workforce; workers and managers can recruit others. Private/closed crowds are not supported.

{\bf Demographics \& Worker Identities.} Previously their workforce was primarily drawn from developing nations (e.g., India, Jamaica~\cite{MWJamaica}, Pakistan, etc.), but they no longer focus exclusively on the developing world~\cite{mw-kulkarni13}. 
Demographic information is not available. 

{\bf Qualifications \& Reputation.} Workers' native languages are recorded. Workers are awarded badges upon completing certifications in order to earn access to certification-restricted tasks, such as tagging of buying guides, image quality, iOS device, Google Drive skills, Photoshop, etc. Requesters select required skills from a fixed list when posting tasks. Each worker's efficiency is measured by an accuracy score, a ranking, and the number of completed tasks. 

{\bf Task Assignment \& Recommendations}. Given worker certifications and accuracy scores, an algorithm dynamically routes tasks accordingly. Workers can select tasks, but the task list on their pages are determined by their qualifications and badges. Task priority is also considered in routing work in order to reduce latency and prevent starvation. Task starvation is reported to have never occurred~\cite{Kulkarni12}. 

{\bf Hierarchy \& Collaboration.} The best workers 
are promoted to managerial positions. Experts check the work of other potential experts in a peer review system. Experts also recruit new workers, evaluate potential problems with Requester-defined tasks, and resolve disagreements in worker responses~\cite{Kulkarni12}.
Provided tools enable worker-to-worker and manager-to-worker communication within and outside of the platform. A worker chat interface allows worker collaboration and discussion while working on tasks. 

{\bf Incentive Mechanisms.} Payments are made via PayPal, MoneyBookers or oDesk. The price per task is set automatically based on estimated completion time, required worker effort and hourly wages in each worker's local time zone. If observed work times vary significantly from estimates, 
the Requester is notified and asked to approve the new price before work continues~\cite{Kulkarni12}.
To determine payout for a certain type of task, they divide a target hourly wage by the average amount of observed time required to complete the task, excluding outliers. This pay structure incentivizes talented workers to work efficiently while ensuring that average workers earn a fair wage. Payout is tiered, with workers whose accuracy is below 80\% only earning 75\% of possible pay. This aims to encourage long-term attention to accuracy \cite{Kulkarni12}.

{\bf Quality Assurance \& Control.} Manual and algorithmic techniques manage quality, including: dynamic work routing, peer management, and social interaction techniques (manager-to-worker, worker-to-worker communication). Worker accuracy is monitored based on majority agreement, and if accuracy falls below a threshold, workers are reassigned to training tasks until they improve. A worker who disagrees with the majority answer can request a manager to review the answer. 
Managers also determine final answers for examples reported as difficult by workers~\cite{Kulkarni12}. 

{\bf Self-service, Enterprise, and API offerings}. Enterprise solutions are supported but self-serve option has been discontinued. They argue that it is the wrong approach to have users design their own tasks and communicate directly with workers in short-duration tasks. Self-serve has been thus replaced by a ``virtual assistant service'' called {\em Premier} that provides help with small projects through post-by-email, an expert finding system, and an accuracy guarantee~\cite{mw-kulkarni13}. 
API offerings: RESTful API; libraries for Python and PHP; sandbox (testing) and production environments; supports iterative, parallel (default), survey, and manual workflows; sends and receives data as JSON; supports optional parameters for filtering workers (blocked, location, age min, age max, gender). 

{\bf Specialized \& Complex Task Support}. Specialized tasks include digitization, categorization, research, feedback, tagging, and others. 
As mentioned above, their API supports various workflows including parallel (same tasks assigned to multiple workers), iterative (workers build on each others' answers), survey, and manual. Some of the specialized tasks supported by the API include processing natural-language responses to user queries, processing images, text, language, speech, or documents processing, creation and organization of datasets, testing, and labeling or dataset classification. A small population of workers performs task on mobiles.


{\bf Ethics \& Sustainability}. Their social mission is to employ marginalized populations of developing nations. This is strengthened by having workers recruit other workers. 
Their pricing structure ensures workers earn fair or above-market hourly wages in their local regions. 
Workers can be promoted to become managers, supporting economic mobility. Managers are further entrusted with the responsibility of hiring new workers, training them, resolving issues, peer review, etc., offering more meaningful and challenging work.

\subsection{oDesk} 

{\bf Distinguishing Features.} Online contractor marketplace; support for complex tasks; flexible negotiated pay model (hourly vs. project-based contracts); work-in-progress screenshots, time sheets, and daily log; rich communication and mentoring support; 
public worker profiles include qualifications, work histories, past client feedback, test scores; team room; payroll/health benefits support~\cite{Ipeirotis-odesk}.

{\bf Whose Crowd?} Have their own workforce. Requesters can bring private workforce: ``Bring your own contractor''.

{\bf Demographics \& Worker Identities.} Global Workforce; public profiles provide name, picture, location, skills, education, past jobs, tests taken, hourly pay rates, feedback, and ratings. Workers' and requesters' identities are verified. 

{\bf Qualifications \& Reputation.} Support is offered for virtual interviews or chat with workers. Work histories, past client feedback, test scores, and portfolios characterize workers' qualifications and capabilities. Workers can also take tests to build credibility. English proficiency is self reported, and workers can indicate ten areas of interest in their profiles.

{\bf Task Assignment \& Recommendations}. Jobs can be posted via: a) Public post, visible to all workers; b) Private invite, directly contacting worker whose skills match the project; and c) both public post with private invites. Ipeirotis discusses research on matching oDesk Requesters to workers~\cite{ipeirotis-odesk-slides}.

{\bf Hierarchy \& Collaboration.} 
Rich communication and information exchange is supported between workers and requesters about ongoing work. A requester can build a team by hiring individual workers to work on the same project, where he may assign team management task to one of the workers. Requesters can monitor their teams using their {\em Team Room} application. When team members login, their latest screenshots are shown, along with their memos and activity meters, which the Requester can see as well. Requesters, managers and contractors can use the message center to communicate and the team application to chat with team members. An online forum 
 and Facebook page
 exist. 

{\bf Incentive Mechanisms.} Workers set their hourly rates and may agree to work at a fixed rate for a project. Workers with higher ratings are typically paid more. A wide variety of payment methods are supported. 
Workers are charged 10\% of the total amount charged to the client. 
With hourly pay, Requesters pay only for the hours recored in the {\em Work diary}.

{\bf Quality Assurance \& Control.} {\em Work diaries} 
report credentials (eg. computer identification information), screenshots (taken 6 times/hr), screenshot metadata (which may contain worker's personal information), webcam pictures (if present/enabled by the worker), and number of keyboard and mouse events. No in-platform QC is provided. In the enterprise solution, QA is achieved by testing, certifications, training, work history and feedback ratings. 

{\bf Self-service, Enterprise, and API offerings}. oDesk offers self-service and enterprise
, with numerous RESTful APIs: Authentication, Search providers, Provider profile, Team (get team rooms), Work Diary, Snapshot, oDesk Tasks API, Job Search API, Time Reports API, Message Center API, Financial Reporting API, Custom Payments API, Hiring, and Organization. API libraries are provided in PHP, Ruby, Ruby on Rails, Java, Python, and Lisp, supporting XML and JSON. 
Ipeirotis reports how the API can be used for microtasks~\cite{ipeirotis-odesk-api}.


{\bf Specialized \& Complex Task Support}. Arbitrarily complex tasks are supported, e.g.\ web development, software development, networking \& information systems, writing \& translation, administrative support, design \& multimedia, customer service, sales \& marketing, business services. Enterprise support is offered for tasks such as writing, data entry \& research, content moderation, translation \& localization, software development, customer service, and custom solutions. 


{\bf Ethics \& Sustainability}. The payroll system lets US and Canada-based workers who work at least 30 hours/wk obtain benefits including: simplified taxes, group health insurance, 401(k) retirement savings plans, and unemployment benefits. Requesters benefit by the platform providing an alternative to staffing firms for payroll and IRS compliance. 

\end{document}